# ROMA (Rank-Ordered Multifractal Analysis) for intermittent fluctuations with global crossover behavior

Sunny W. Y. Tam[1,2], Tom Chang[3], Paul M. Kintner[4], and Eric M. Klatt[5]

[1]*Plasma and Space Science Center, National Cheng Kung University, Tainan 70701, Taiwan*
[2]*Institute of Space, Astrophysical and Plasma Sciences, National Cheng Kung University, Tainan 70101, Taiwan*
[3]*Kavli Institute for Astrophysics and Space Research, Massachusetts Institute of Technology, Cambridge, MA 02139, USA*
[4]*School of Electrical and Computer Engineering, Cornell University, Ithaca, NY 14850, USA*
[5]*Applied Physics Laboratory, Johns Hopkins University, Laurel, MD 20723, USA*

## ABSTRACT

Rank-Ordered Multifractal Analysis (ROMA), a recently developed technique that combines the ideas of parametric rank ordering and one parameter scaling of monofractals, has the capabilities of deciphering the multifractal characteristics of intermittent fluctuations. The method allows one to understand the multifractal properties through rank-ordered scaling or non-scaling parametric variables. The idea of the ROMA technique is applied to analyze the multifractal characteristics of the auroral zone electric field fluctuations observed by SIERRA. The observed fluctuations span across contiguous multiple regimes of scales with different multifractal characteristics. We extend the ROMA technique such that it can take into account the crossover behavior -- with the possibility of collapsing probability distributions functions (PDFs) -- over these contiguous regimes.

## I. INTRODUCTION

Increasing evidence based on the analyses of satellite and spacecraft measurements has shown that plasma fields in space are often intermittent in nature [1-6]. Associated with intermittency, these field fluctuations exhibit multifractal characteristics. The fluctuation series of these fields have popularly been analyzed with structure functions and/or partition functions [6-10]. These traditional methods of multifractal analysis determine the fractal properties of the fluctuations under various moment orders generally based on the entire set of observed or simulated fluctuations. If such moment orders demonstrate self-similar fractal properties, then the fluctuations exhibit monofractal nature. In this case, the probability distribution functions (PDFs) of the magnitude of the fluctuations at different scales can be mapped onto one master PDF in terms of one parameter with a single scaling (fractal) power. More often, at least for fluctuations associated with space plasmas, different fractal properties are found with different moment orders, meaning that the fluctuations have multifractal characteristics. For such fluctuations, the idea of one single scaling power can no longer be applied to relate the PDFs at different time scales.

Recently, a new technique [11] for analyzing intermittent fluctuations has been developed to describe parametrically the explicit multifractal characteristics and indicate how they are distributed within the fluctuations. The technique, known as Rank-Ordered Multifractal Analysis (ROMA), retains the spirit of the traditional structure function analysis and combines it with the idea of one-parameter scaling of monofractals. It was first applied to the results of a large-scale two-dimensional (2-D) magnetohydrodynamic (MHD) turbulence simulation.

We aim to apply the idea of ROMA to intermittent electric field fluctuations measured in the auroral zone. Unlike the 2-D MHD fluctuations, the auroral zone electric field fluctuations sometimes exhibit crossover behavior spanning across contiguous regimes of scales with different multifractal characteristics. Using the auroral zone electric field measured by the SIERRA (Sounding of the Ion Energization Region: Resolving Ambiguities) sounding rocket as an example, we shall apply the ROMA technique for fluctuations exhibiting crossover behavior straddling such contiguous regimes of scales.

This paper is structured as follows: In Section II, we apply the traditional structure function analysis to a fluctuating series of an electric field measured by the SIERRA sounding rocket. Based on the results of the analysis of the fluctuations, different regimes of scales of multifractal characteristics are identified. In Section III, ROMA is applied to each regime. Using the contiguous scaling property over the rank-ordered scales, we then extend the technique of ROMA in Section IV, in which we shall show how a global parametric variable may provide the scaling over all regimes of scales.

## II. STRUCTURE FUNCTION ANALYSIS AND SINGLE FRACTAL POWER SCALING

The data used in this study are a time series of an electric field component perpendicular to the magnetic field, as measured by the SIERRA sounding rocket in the auroral zone. The time series of the fluctuations, as shown in Fig. 1 (top), corresponds to where the rocket was between 550 km altitude and its apogee at 735 km altitude. The averaged spectral density is broadband as shown in Fig. 1 (bottom). It has been suggested by Chang [12] that such broadband signature might be the manifestation of intermittent turbulence. In fact, fluctuations in a subset of this time series have been shown to be intermittent based on analyses with the techniques of Probability Distribution Function (PDF), wavelet analysis and local intermittency measure, indicating the fluctuations are multifractal in nature [5].

The origin of intermittent fluctuations in magnetized plasmas was interpreted ([13] and references contained therein) as the result of the sporadic mixing and/or interactions of localized pseudo-coherent structures. The dominant forms of such structures in the auroral zone probably include those similar to the nearly 2D oblique potential structures simulated by Seyler [14] based on the reduced MHD formulation of the inertial Alfvén fluid equations and also other small scale kinetic coherent structures. The interactions of these structures are the manifestation of localized reconnections mediated by intermittent inertial Alfvén waves and other electromagnetic and electrostatic waves, generalized resistivity and/or coarse-grained dissipation. Thus, we expect a significant fraction of such fluctuations to be electrostatic and transverse [12], perhaps interspersed with small components of electrostatic [15] and/or electromagnetic waves. When detected in the spacecraft frame, the signatures of the interacting and nearly electrostatic structures are Doppler shifted [16-18]. They have been shown to have relatively slow moving speeds in the rest frame [19]. Thus, a dominant fraction of the fluctuations may be recognized as the spatial fluctuations of low frequency intermittent inertial Alfvén turbulence and small scale kinetic turbulence. And the time series and time scales may be interpreted mainly in terms of spatial scales and spatial fluctuations. Assuming the horizontal speed $U$ of the spacecraft is much larger than that of the movements of the broadband turbulent fluctuations and the geomagnetic field is essentially vertical, the time scales $\tau$ to be discussed below may be viewed approximately as spatial scales $\Delta \approx U\tau$, where $U \approx 1.5$ km/s. Nevertheless, it must be recognized that entrained within such observed broadband turbulence there may be small fractions of electromagnetic and Doppler shifted electrostatic propagating waves.

We shall first apply the structure function analysis, a traditional method of multifractal analysis, to the time series of the electric field, $E$. For a given time scale $\tau$, we calculate the increment of the field $\delta E \equiv E(t+\tau) - E(t)$. We can then obtain $P(|\delta E|, \tau)$, the PDFs of the absolute value of $\delta E$ at scale $\tau$, as shown in Fig. 2, which include the plots for all the time scales (from 5 to 1280 ms) considered in our study. Note that the PDFs are normalized such that:

$$\int_0^\infty P\left(|\delta E|, \tau\right) d|\delta E| = 1 . \quad (1)$$

The structure function at time scale $\tau$ with moment order $q$ is defined to be:

$$S_q(\tau) \equiv \left\langle |\delta E(\tau)|^q \right\rangle = \int_0^\infty |\delta E|^q P\left(|\delta E|, \tau\right) d|\delta E| , \quad (2)$$

where $\langle \ldots \rangle$ denotes averaging over *t*. In Eq. (2), the moment order *q* is taken to be non-negative to ensure that the right-hand side of the equation does not diverge. With non-negative *q*, one then looks for the scaling behavior

$$S_q(\tau) \sim \tau^{\zeta_q} . \quad (3)$$

Generally, the "fractal dimension" $\zeta_q$ may not vary linearly with *q*. But if $\zeta_q$ is linearly proportional to *q*, then the fractal properties of all the moments can be characterized by a single fractal number. That would be the case for monofractals since the fractal characteristics of all moment orders are similar to each other. This, in turn, renders a constant value for the "Hurst exponent" defined as:

$$H(q) \equiv \zeta_q / q . \quad (4)$$

For multifractals, the Hurst exponent would vary with *q*. In the case of monofractals, one may scale the PDFs for different $\tau$'s with one scaling function $P_s$ and one power-law parameter in terms of a constant scale exponent *s* as follows:

$$P(|\delta E|, \tau) = \left(\tau/\tau_0\right)^{-s} P_s\left( |\delta E| (\tau/\tau_0)^{-s} \right) , \quad (5)$$

where $\tau_0$ is a reference time scale. Such scaling implies a functional relation between the power-law scale invariants [20]:

$$P(|\delta E|,\tau)\left(\tau/\tau_0\right)^s \text{ and } Y = |\delta E|(\tau/\tau_0)^{-s}, \tag{6}$$

with a constant value of $s$ and has been employed in studies of various kinds of fluctuating fields, including the stock market indices [21], magnetic fluctuations in space [13, 22] and fluctuating events of other natural or experimental systems [23].

Figure 3 shows the plots of $S_q(\tau)$ vs. $\tau$ at a few moment orders $q$ for the electric field fluctuations. It is clear from the plot for $q=1$ (top left panel) that Eq. (3) cannot be applied throughout the entire range of time scales. There are at least two regimes in $\tau$ where Eq. (3) may be separately applied, with the crossover time scale appearing at $\tau \approx 80$ ms. Information based on the other panels of the figure indicates that the time scales above 80 ms should be further divided into different regimes with $\tau \approx 160, 320$ ms being other crossover scales. Thus, the auroral zone electric field fluctuations seem to exhibit four different regimes over the time scales up to 1280 ms: Regime 1 from $\tau \approx 5$ to 80 ms; Regime 2 from $\tau \approx 80$ to 160 ms; Regime 3 from $\tau \approx 160$ to 320 ms; and Regime 4 for $\tau \gtrsim 320$ ms. We shall rank order the time regimes parametrically by an index $i = 1, 2, 3$ and 4, and study the multifractal characteristics of each regime separately.

It remains to be determined, however, whether the fluctuations would exhibit monofractal behavior within each of the rank-ordered regimes. We shall continue with the discussion on structure function analysis in detail only for the case of Regime 1. The dotted lines in Fig. 3 are straight lines fitted to the points that belong to Regime 1. Their slopes are the corresponding values for $\zeta_q$, which are shown in Fig. 4 along with the Hurst exponent $H(q)$. In the top panel of the figure, the plot of $\zeta_q$ vs. $q$ is nearly --- but not quite --- a straight line through the origin, suggesting that the fluctuations are multifractal, but perhaps close to monofractal within the time scales of Regime 1. The multifractal nature of the fluctuations, however, is more clearly indicated by the non-constant value of $H(q)$ in the bottom panel. Therefore, we expect that the scaling relation in the form of Eq. (5) would not hold even within this regime. If the scaling relation were to hold, one would be able to find a constant value of *s* that, based on Eq. (5), enables the PDFs at all five time scales of Regime 1 to collapse into a single function $P_s(Y)$ where *Y* is given by Eq. (6). Using $\tau_0 = 5$ ms and trying a wide range of values for *s*, we map each $P(|\delta E|, \tau)$ to its respective scaling function $P_s(Y)$, and find that the closest agreement among the five time scales occurs at $s = 0.69$, for which the results are shown in the top panel of Figure 5. Although the PDFs collapse very well to a single curve for fluctuations at small magnitudes, up to about $Y = 0.5$, their agreement is not good at all toward the tail of the distributions. As we try to choose a different value of *s* to improve the agreement at the tail, such as the results shown in the bottom panel of Figure 5 with $s = 0.9$, the mapped PDF do not collapse onto each other at small *Y* any more. Thus, the fluctuations indeed do not obey the scaling relation (5), a further indication of their non-monofractal or multifractal nature.

## III. APPLICATION OF ROMA TECHNIQUE TO INDIVIDUAL REGIMES

The ROMA technique can be applied to each of the four rank-ordered regimes for the auroral zone electric field fluctuations. In this section, we shall demonstrate the method and its application in detail to time scales of Regime 1 only. We have seen in the previous section that because the fluctuations are not monofractal in nature, for any given constant value of the scaling exponent *s*, the PDFs at different time scales collapse onto a single curve, at best, over only a portion in the domain of $Y$. The idea of ROMA is to divide the domain of $Y$ into separate ranges, and for each range, to determine a value of $s$ that would satisfy Eq. (5). As a result, the scaling exponent $s$ can take on different values over different ranges of $Y$. To solve for the scaling exponent for a given range of $Y = [Y_{\mathrm{low}},\ Y_{\mathrm{high}}]$, we construct a range-limited structure function $S'_q(\tau)$ with prescribed trial values of $s$:

$$S'_q(\tau) = \int_{Y_{low}(\tau/\tau_0)^s}^{Y_{high}(\tau/\tau_0)^s} \left|\delta E(\tau)\right|^q P\left(\left|\delta E\right|, \tau\right)\, d\left|\delta E\right| . \tag{7}$$

Note that unlike the traditional structure function analysis, ROMA can be carried out with negative values of $q$ without the problem of $S'_q(\tau)$ diverging, except for the range that includes $Y_{\mathrm{low}} = 0$. Similar to the approach with the traditional structure function method, we then look for the scaling behavior

$$S'_q(\tau) \sim \tau^{\zeta'_q} , \tag{8}$$

such that

$$\zeta'_q = qs . \tag{9}$$

If a (unique) solution exists for $s$ for the chosen range of $Y$, then the fractal behavior of the fluctuations within this range is monofractal and characterized by the fractal number (local Hurst exponent), $s$.

To demonstrate the technique for determining $s$, Figure 6 shows the results of the application of ROMA to the range of $Y = [0.8,\ 1.2]$ for the electric field fluctuations at time scales of Regime 1, using $q = 2$ as an example. In the figure and in the following discussion, the subscript “1” is added to “$s$” and “$Y$” to denote that the application is for the $i = 1$ regime only. Every point plotted in the figure is obtained by finding $\zeta'_q$ with Eqs. (7) and (8) based on a given value of $s_1$ and a fixed $q$. The dotted lines correspond to Eq. (9), with $s$ replaced by $s_1$.

Because the scaling exponent must satisfy Eq. (9), the solution for $s_1$ has to fall on the dotted line. The plot in the top panel of Fig. 6 indicates that the solution for $s_1$ is approximately 0.8. We increase the resolution in our trial values of $s_1$ and find that a more precise solution for the scaling parameter to be $s_1 \simeq 0.804$. We then check the validity of this solution by using this value of $s_1$ to obtain $\zeta'_q$ for various values of *q*; a valid solution should yield result that agrees with Eq. (9) (with *s* replaced by $s_1$) for every *q*. Figure 7 confirms the validity of our solution, as every point in the plot of $\zeta'_q$ vs. *q* falls on the dotted line that corresponds to Eq. (9).

Similarly, we find values for the scaling exponent $s_1$ for other ranges of $Y_1$. Figure 8 shows the solutions we get from all the ranges, with each horizontal line in the plot indicating the value of $s_1$ for the corresponding $Y_1$-range. The variation of $s_1$ for different ranges of $Y_1$ indicates the multifractal behavior of the fluctuations within the time scales of Regime 1. This is the approximate multifractal spectrum for Regime 1 using the ROMA technique. The spectrum $s_1(Y_1)$ indicates how the various fractal properties of the fluctuations in Regime 1 are distributed. The spectrum is implicit since $Y_1$ depends on $s_1$. In principle, one may apply this technique using smaller and smaller $Y_1$-ranges and get better and better resolution in $Y_1$. By doing so, one can visualize $s_1$ to be a continuous function of $Y_1$; a plot of $s_1(Y_1)$ would appear as a continuous spectrum. However, for practical purpose, such continuity usually cannot be achieved for a finite time series of data. As the size of the $Y_1$-ranges keeps decreasing, fewer and fewer data points are available for each range; the validity of the statistics used to determine $S'_q(\tau)$ will result in unreliable fluctuations.

The scaling for the range that includes the smallest $Y_1$, namely $Y_1 = [0,\ 0.1]$ in our calculations, is found to be characterized by $s_1 = 0.694$, as shown in Fig. 8. Such a result for $s_1$ is very close to 0.69, a value that would lead to the collapse of the PDFs in the corresponding range of $Y_1$ for the five time scales in Regime 1, as indicated in the top panel of Fig. 5. As $Y_1$ gradually increases, we find that $s_1$ initially decreases to about 0.66 for $Y_1 = [0.1,\ 0.3]$, but increases rapidly after that to above 0.85 when $Y_1 = [1.2,\ 1.8]$. Beyond such a range in $Y_1$, $s_1$ begins to show a slightly decreasing trend, maintaining a value as high as 0.83 when $Y_1$ increases to the range 2.6 – 3.6. One may notice the resemblance between the general shape of $s_1(Y_1)$ in Fig. 8 and $H(q)$ in the bottom panel of Fig. 4, which may be understood as follows: From Eqs. (3) and (4), it is clear that for a given moment order *q*, the Hurst exponent depends on how $S_q(\tau)$ scales with $\tau$ for the time scales within the same regime. The structure function $S_q(\tau)$ can be understood as a sum of contribution from all parts of the PDF $P(|\delta E|, \tau)$, as suggested by Eq. (2). For a given PDF, there is a certain portion of $|\delta E|$ that would contribute most to the structure function, and the value of such $|\delta E|$ would depend on *q*. Generally, as *q* increases, one would find the portion of $|\delta E|$ with the most significant contribution moving toward the tail of the PDF. If we express the PDF as a function of $Y_1$ rather than the unscaled variable $|\delta E|$, it would mean that the range that contributes most to the structure function would be found at a larger value of $Y_1$ as $q$ increases. If we use only such a range in $Y_1$, as opposed to any other similar partial ranges, to construct the partial structure functions $S'_q(\tau)$ in Eq. (7), the resulting $\zeta'_q$ based on Eq. (8), in principle, should be the most reasonable approximation for the fractal dimension $\zeta_q$ (obtained from the full structure functions by Eq. (3)) at the moment order $q$ where such a $Y_1$-range dominates. Thus, to a certain extent, the variation of $\zeta'_q$ through ranges in $Y_1$ should characterize the qualitative behavior of $\zeta_q$ vs. $q$. If we now compare Eq. (9) with Eq. (4), which can be written as $\zeta_q = qH$, we can see that the relationship between $s_1(Y_1)$ and $H(q)$ is the same as that between $\zeta'_q$ and $\zeta_q$; $s_1(Y_1)$ characterizes the qualitatively behavior of $H(q)$ to a certain extent.

Generally, how good the resemblance between the shape of $s(Y)$ and $H(q)$ depends on how well the domains of $|\delta E|$ and $Y$ are correlated. In particular, the above argument for the resemblance does not apply well when a narrow range in the domain of $|\delta E|$ corresponds to a wide range in the domain of $Y$. Such a situation may occur --- as one may infer from Eq. (11) below --- when there is a significant decrease in $s$ over a small range in $Y$. In that case, the Hurst exponent at the moment order that emphasizes the narrow range in $|\delta E|$ would characterize the average fractal behavior of a wide range in $Y$. Because this average behavior is not detailed enough to describe the change in the scaling behavior over such a range of $Y$, the variation of $H(q)$ would fail to reflect the specific variation or fluctuation of the rank-ordered spectrum $s(Y)$, which is a more accurate description of the local fractal behavior. Therefore, when $H(q)$ and $s(Y)$ are found to have very different shapes, one may consider that as an indication of inadequacy of the traditional structure function analysis. Later, we shall see that such a situation actually occurs for the auroral electric field fluctuations in one of the regimes.

From the above discussion of the relationship between $s_1(Y_1)$ and $H(q)$, one perceives that each value of $s_1$ plays the role of the Hurst exponent for a small local range of the PDFs. Thus, when local ranges of the PDFs for the fluctuations are considered, we expect $s_1(Y_1)$ to share the utilities of the Hurst exponent in analyzing fluctuations. For instance, based on the results that $s_1(Y_1)$ is overall considerably larger than 0.5, the classical demarcation for the Hurst exponent between persistency and anti-persistency, we may wish to conclude that the fluctuations are all persistent at the time scales of Regime 1. However, the apparent persistency may be due to kinetic effects, which are probably important in this regime of small scales. In addition, at small values of $Y_1$, which correspond to small sizes of $|\delta E|$, the scaling exponent $s_1$ increased rapidly, an indication of possible developing instability and turbulence. The fluctuations seem to settle down to a more stable and developed turbulent state as $Y_1$ becomes larger, as the values of $s_1$ seem to become more and more slowly varying.

With the determination of the rank-ordered spectrum of the scaling exponent $s_1(Y_1)$, the scaling relationship, instead of Eq. (5), can now be expressed as

$$P_1(|\delta E|,\tau) = \left(\tau/\tau_0\right)^{-s_1(Y_1)} P_{s1}\left( |\delta E|(\tau/\tau_0)^{-s_1(Y_1)} \right), \qquad (10)$$

corresponding to a second rank-ordered parameter

$$Y_1 = |\delta E|(\tau/\tau_0)^{-s_1(Y_1)} \;, \qquad (11)$$

where $P_1\left(|\delta E|,\tau\right)$ denotes $P\left(|\delta E|,\tau\right)$ for $\tau$ belonging to Regime 1. We emphasize that, unlike that for monofractal scaling, fluctuations within Regime 1 do not form a single power-law scale-invariant with a constant value of $s_1$. Instead, the fluctuations are grouped into a spectrum of power-law scale-invariants with different fractal powers characterized by $s_1(Y_1)$. "$Y_1$" is, nevertheless, a parametric power-law scaling variable within Regime 1. In principle, Eq. (10) is valid for the entire range of $|\delta E|$ at any time scale $\tau$ within Regime 1. In practice, however, the scaling relationship holds only up to a finite value of $|\delta E|$, above which the number of samples from the data is not enough to give accurate statistical results for $S'_q(\tau)$. Correspondingly, since $Y_1$ is affected by such a limitation related to statistics, it is expected that the accuracy of the results associated with power-law scaling can only be maintained up to a certain finite value of $Y_1$. In the domain where there is no such problem due to statistics, ROMA, when applicable, has a distinct advantage over the traditional structure function analysis in that the new technique is able to provide additional information about the multifractal fluctuations. First, note that with $s_1(Y_1)$ determined, one is able to solve for $Y_1$ in the implicit equation (11) for given $|\delta E|$ and $\tau$. Using the solution for $Y_1$, one is then able to find the corresponding scaling exponent $s_1$, and, based on Eq. (9) (with $s$ now replaced by $s_1$), $\zeta'_q$ as well. In other words, given $|\delta E|$ and $\tau$, ROMA is able to provide the fractal properties associated with fluctuations of that magnitude at that time scale within the parametric rank $i=1$. In contrast, $\zeta_q$, which is determined from Eq. (3) based on the traditional structure function analysis, reflects only the fractal property of the entire set of observed fluctuations without distinguishing between the various magnitudes and time scales. Second, ROMA provides a second rank-ordered parameter $Y_1$, which, as discussed above, is solvable for given $|\delta E|$ and $\tau$.

We may now map the PDF of the electric field fluctuations at each of the time scales in Regime 1 to a scaling function $P_{s1}(Y_1)$, based on the following equation derived from Eqs. (10) and (11):

$$P_{s1}(Y_1) = (\tau/\tau_0)^{s_1(Y_1)} P_1(|\delta E|, \tau\ ), \qquad (12)$$

using $\tau_0 = 5$ ms and the profile of $s_1(Y_1)$ from linear interpolation through the midpoints of the ranges in $Y_1$. Figure 9 shows the results of the mapping for the various $\tau$ . One can see that the results for the various $P_{s1}(Y_1)$ agree quite well up to about $Y_1 = 1.8$ . Beyond that, there is increased discrepancy due to the aforementioned problem associated with poor statistics. Nevertheless, at those relatively large values of $Y_1$, the various $P_{s1}(Y_1)$ still appear to agree better than those obtained by the traditional single fractal power scaling (c.f. Fig. 5).

Similar calculations may be carried out for the regimes $i = 2$, 3 and 4, yielding $s_i(Y_i)$ and $P_{si}(Y_i)$ as shown in Figs. 10-15. The agreement between the scaled PDFs for the time scales within the same regime verifies the validity of the results of $s_i(Y_i)$, at least in the ranges of $Y_i$ where the samples are sufficient for the statistics to be meaningful. We shall briefly discuss the variations and the probable meaning of the rank-ordered spectrum for each regime. For Regime 2, the shape of $s_2(Y_2)$ (Fig. 10) is generally comparable to the variation of the corresponding Hurst exponent $H(q)$ found for this regime, as shown in the top panel of Fig. 16, due to a similar reason as in the case of Regime 1. At small values of $Y_2$, $s_2$ exhibits fluctuations in the range around 0.5. Such fluctuating behavior is rather similar to that of $s_1$ at small values of $Y_1$ (Fig. 8), except that the values for $s_2$ are considerably lower. Thus, the developing turbulence seems to be of a mixture of persistent and anti-persistent nature, probably as a result of effects beyond the kinetic range starting to play a non-negligible role at the scales of this regime. As $Y_2$ becomes larger, the values of $s_2$ become more stable, indicative of the turbulence settling down to a more stable and developed state, similar to the case for Regime 1. The apparent persistent nature of the fluctuations suggested by the values of $s_2$ at large $Y_2$ is perhaps due to kinetic effects still being more dominant than those of larger scales.

Regime 3, in contrast to the previous two regimes, features a rank-ordered spectrum $s_3(Y_3)$ whose shape does not resemble that of the corresponding $H(q)$ (Fig. 12 and the middle panel of Fig. 16). This can be explained, based on our discussion earlier, by the significant decrease in the value of $s_3$, which drops from 0.677 to 0.285 as $Y_3$ increases from [5, 8] to [14, 19]. In addition to such a relatively wide range of values covered by the rank-ordered spectrum in this regime as compared with Regimes 1 and 2, $s_3(Y_3)$ also fluctuates over a large range in $Y_3$ rather than only near small values of $Y_3$, suggesting that the turbulence at the scales of this regime is highly unstable. In fact, one may infer about the unstable turbulent state just by qualitatively comparing the PDFs $P(|\delta E|, \tau)$ at the two time scales considered in this regime, $\tau = 160$ and 320 ms. As shown in Fig. 17, when $|\delta E|$ is small, the plots of the PDFs at the two time scales are close to each other. But the separation between the plots becomes wider when $|\delta E|$ increases to about 5 to 8 mV/m for $\tau = 160$ ms (indicated by the thick arrow in the figure), and then becomes narrow again as $|\delta E|$ increases to about 14 to 19 mV/m (indicated by the thin arrow in the figure), and wider again when $|\delta E|$ increases further. The highly fluctuating separation between the two plots means that the shape of the two PDFs is far from self-similar, an indication that the turbulent fluctuations are in an unstable state. The change in this separation as $|\delta E|$ varies is also associated with the change in the local scaling exponent $s_3$: the wider this separation, the larger $s_3$; thus giving rise to the fluctuations in the rank-ordered spectrum $s_3(Y_3)$ (Fig. 12).

For Regime 4, the rank-ordered spectrum $s_4(Y_4)$ shares the same general shape as the corresponding $H(q)$ (Fig. 14 and the bottom panel of Fig. 16). At small values of $Y_4$, there is an increase in $s_4$. After reaching a peak value close to 0.5, $s_4$ then decreases monotonically (at least in the range where the statistics are reliable) as $Y_4$ increases further. The monotonically decreasing trend of the rank-ordered spectrum, together with the anti-persistent nature of the fluctuations, seem to be the signature of developing inertial Alfvén MHD turbulence or even classical MHD turbulence, as such qualitative features were also found in the ROMA calculations that analyzed the results of a two-dimensional MHD simulation [11].

## IV. APPLICATION OF ROMA ACROSS REGIMES OF TIME SCALES

We have demonstrated the utilities of ROMA on fluctuations within different regimes of time scales in terms of two rank-ordered parametric variables, the "index $i$" and the "power-law scaling variable, $Y_i$". This idea can be applied to any multi-parameter rank-ordered regions such as those characterizing anisotropy, inhomogeneity and unsteadiness. The indexing of rank ordering does not need to follow the size of the rank-ordered parameter(s). In fact, size or numerical value may not be the criterion for rank ordering. When the rank ordering is in fact size dependent and contiguous, a further extension of the ROMA technique may be applied. We shall now describe the global crossover behavior over contiguous rank-ordered regimes using the example of the electric field fluctuations in the auroral zone.

The 4 ROMA spectra are applicable to the 4 separate time regimes separately. There is also a roughly defined common time scale between any two adjacent regimes: $\tau \approx 80$ ms belonging to both Regimes 1 and 2, $\tau \approx 160$ ms belonging to both Regimes 2 and 3, and $\tau \approx 320$ ms being common to Regimes 3 and 4. Assuming the non-power law crossover ranges between contiguous time regimes are very narrow such that the changes across regimes are essentially characterized by abruptly changing but piecewise continuous power-laws, we may extend the ROMA technique to determine an implicit "multi-power" global scaling variable associated with the fluctuations covering all four regimes of time scales as follows: ROMA is applicable to the *i*-th regime (*i* goes from 1 to 4 in our case), and $P_i(|\delta E|,\tau)$ with time scale $\tau$ that belongs to that regime can be described by the scaling relationship.

$$P_i(|\delta E|,\tau) = \left(\tau/\tilde{\tau}_i\right)^{-s_i(Y_i)} P_{si}\left(|\delta E|\left(\tau/\tilde{\tau}_i\right)^{-s_i(Y_i)}\right), \tag{13}$$

where $\tilde{\tau}_i$ is the smallest time scale of the *i*-th regime, and $s_i(Y_i)$ is the rank-ordered spectrum of scaling parameter for the regime, with $Y_i$ being the "scaled" parametric scaling variable implicitly provided by the equation:

$$Y_i = |\delta E|\left(\tau/\tilde{\tau}_i\right)^{-s_i(Y_i)}. \tag{14}$$

The four scaling variables as well as the scaled PDFs for $i = 1,2,3,4$ are related due to the assumed piecewise continuous property across the contiguous regimes. Straightforward algebra leads to the following recursion relations:

$$Y_{i+1} = Y_i\left(\tilde{\tau}_{i+1}/\tilde{\tau}_i\right)^{s_i(Y_i)}. \tag{15}$$

$$P_{s(i+1)}\left(|\delta E|\right) = \left(\tilde{\tau}_{i+1}/\tilde{\tau}_i\right)^{-s_i(Y_i)} P_{si}\left(|\delta E|\left(\tilde{\tau}_{i+1}/\tilde{\tau}_i\right)^{-s_i(Y_i)}\right). \tag{16}$$

Applying this idea to the four time regimes, we find a global scaling variable $Y_{global}$ across the four regimes with $P_{s1}$ being the global scaling function as follows:

$$Y_{global} \equiv \begin{cases} |\delta E|\left(\tau/\tilde{\tau}_1\right)^{-s_1(Y_1)} = Y_1 & \text{Regime 1} \\ |\delta E|\left(\tau/\tilde{\tau}_2\right)^{-s_2(Y_2)}\left(\tilde{\tau}_2/\tilde{\tau}_1\right)^{-s_1(Y_1)} = Y_2\left(\tilde{\tau}_2/\tilde{\tau}_1\right)^{-s_1(Y_1)} & \text{Regime 2} \\ |\delta E|\left(\tau/\tilde{\tau}_3\right)^{-s_3(Y_3)}\left(\tilde{\tau}_3/\tilde{\tau}_2\right)^{-s_2(Y_2)}\left(\tilde{\tau}_2/\tilde{\tau}_1\right)^{-s_1(Y_1)} & \\ \quad = Y_3\left(\tilde{\tau}_3/\tilde{\tau}_2\right)^{-s_2(Y_2)}\left(\tilde{\tau}_2/\tilde{\tau}_1\right)^{-s_1(Y_1)} & \text{Regime 3} \\ |\delta E|\left(\tau/\tilde{\tau}_4\right)^{-s_4(Y_4)}\left(\tilde{\tau}_4/\tilde{\tau}_3\right)^{-s_3(Y_3)}\left(\tilde{\tau}_3/\tilde{\tau}_2\right)^{-s_2(Y_2)}\left(\tilde{\tau}_2/\tilde{\tau}_1\right)^{-s_1(Y_1)} & \\ \quad = Y_4\left(\tilde{\tau}_4/\tilde{\tau}_3\right)^{-s_3(Y_3)}\left(\tilde{\tau}_3/\tilde{\tau}_2\right)^{-s_2(Y_2)}\left(\tilde{\tau}_2/\tilde{\tau}_1\right)^{-s_1(Y_1)} & \text{Regime 4} \end{cases} \quad (17)$$

with

$$P(|\delta E|,\tau) = \frac{Y_{global}}{|\delta E|} P_{s1}\left(Y_{global}\right), \quad (18)$$

applicable to all four regimes. Obviously this technique is not limited by the number of regimes of time scales exhibited by the fluctuations provided that they are contiguous.

The bottom panel of Figure 18 shows the profiles of $P_{s1}\left(Y_{global}\right)$ resulting from the mapping of $P(|\delta E|,\tau)$ of all the time scales of the four regimes whose fractal exponents are given in the top panel of the same figure.

To summarize, we have demonstrated that the ROMA technique, when applicable, has advantages over the traditional structure function method in analyzing fluctuations that exhibit multifractal behavior. ROMA is able to provide the specific fractal properties for fluctuations of given magnitude and given time scale, as well as scaling associated with the PDFs of the fluctuations within certain range of time scales. In this study, we applied the ROMA technique to the auroral zone electric field fluctuations with two rank-ordered parameters across contiguous multiple regimes of different physical processes. The transition over the regimes is characterized by a crossover behavior expressed in terms of a global scaling variable and a global scaling function.

ACKNOWLEDGMENTS

This research is partially supported by the National Science Council of R.O.C. under grant NSC 96-2111-M-006-002-MY3 and the AFOSR and NSF of the U.S. government.

## Figure Captions

FIG. 1. (Top) Time series of an electric field component perpendicular to the magnetic field in the auroral zone, as measured by the SIERRA sounding rocket when the rocket was between 550 km and its apogee at 735 km altitude. (Bottom) Average spectral density of the electric field component over the duration.

FIG. 2. PDF $P(|\delta E|, \tau)$ at nine different time scales varying from $\tau = 5$ to 1280 ms. The unit of $|\delta E|$ is mV/m.

FIG. 3. Plots of $S_q(\tau)$ vs. $\tau$ for moment order $q$ of integer values from 1 to 5. The plots indicate the existence of four distinct regimes in time scale: Regime 1: $\tau = 5$ to 80 ms; Regime 2: $\tau = 80$ to 160 ms; Regime 3: $\tau = 160$ to 320 ms; Regime 4: $\tau \geq 320$ ms. The dashed lines indicate fitting through the five time scales of Regime 1.

FIG. 4. (Top) Plot of $\zeta_q$ vs. $q$; (Bottom) Plot of $H(q)$ vs. $q$; results based on the traditional structure function analysis over the five time scales of Regime 1.

FIG. 5. Profiles of $P_s(Y)$ based on applications of single-parameter scaling (Eqs. (5) and (6)) with two different values of $s$ for $\tau$ in Regime 1.

FIG. 6. Plots of $\zeta'_q$ vs. $s_1$ for the determination of the scaling parameter for the range $Y_1 = [0.8, 1.2]$ in the application of ROMA to Regime 1. Moment order $q = 2$ is used as an example here. The dotted line represents the plot of Eq. (9) with $s$ replaced by $s_1$. From the top

panel, the solution is $s_1 = 0.8$. With increased resolution (bottom panel), a more precise solution is found to be $s_1 = 0.804$.

FIG. 7. Plot of $\zeta'_q$ vs. $q$ to confirm the solution $s_1 = 0.804$ for the range $Y_1 = [0.8, 1.2]$ in the application of ROMA to Regime 1. The solution is confirmed as the plot coincides with that of Eq. (9) (with $s$ replaced by $s_1$), which is represented by the dotted line.

FIG. 8. Profile of the rank-ordered spectrum for the scaling exponent of Regime 1, $s_1(Y_1)$. The extent of the horizontal lines indicates the ranges in $Y_1$ over which the scaling exponent $s_1$ is obtained.

FIG. 9. Scaling function $P_{s1}(Y_1)$ obtained from the PDF at each time scale of Regime 1.

FIG. 10. Profile of the rank-ordered spectrum for the scaling exponent of Regime 2, $s_2(Y_2)$. The meaning of the horizontal lines is similar to that in Fig. 8.

FIG. 11. Scaling function $P_{s2}(Y_2)$ obtained from the PDF at each time scale of Regime 2.

FIG. 12. Profile of the rank-ordered spectrum for the scaling exponent of Regime 3, $s_3(Y_3)$. The meaning of the horizontal lines is similar to that in Fig. 8.

FIG. 13. Scaling function $P_{s3}(Y_3)$ obtained from the PDF at each time scale of Regime 3.

FIG. 14. Profile of the rank-ordered spectrum for the scaling exponent of Regime 4, $s_4(Y_4)$. The meaning of the horizontal lines is similar to that in Fig. 8.

FIG. 15. Scaling function $P_{s4}(Y_4)$ obtained from the PDF at each time scale of Regime 3.

FIG. 16. The Hurst exponent $H$ as a function of the moment order $q$ based on the traditional structure function analysis over the time scales of Regime 2 (top panel), Regime 3 (middle panel), and Regime 4 (bottom panel).

FIG. 17. Probability distribution functions $P\left(\left|\delta E\right|,\tau\right)$ for $\tau$ in Regime 3. The unit of $\left|\delta E\right|$ is mV/m. The separation between the two PDFs in the plot varies: wider at the level indicated by the thick arrow and narrower at the level indicated by the thin arrow.

FIG. 18. Top: The rank-ordered spectra $s_1$ (black), $s_2$ (red), $s_3$ (blue), and $s_4$ (green) over the ranges of $Y_{global}$. Plots are obtained based on application of Eq. (17) to the results in Figs. 8, 10, 12 and 14. Bottom: Global scaling function $P_{s1}(Y_{global})$ obtained from the PDFs at all the time scales of the four regimes.

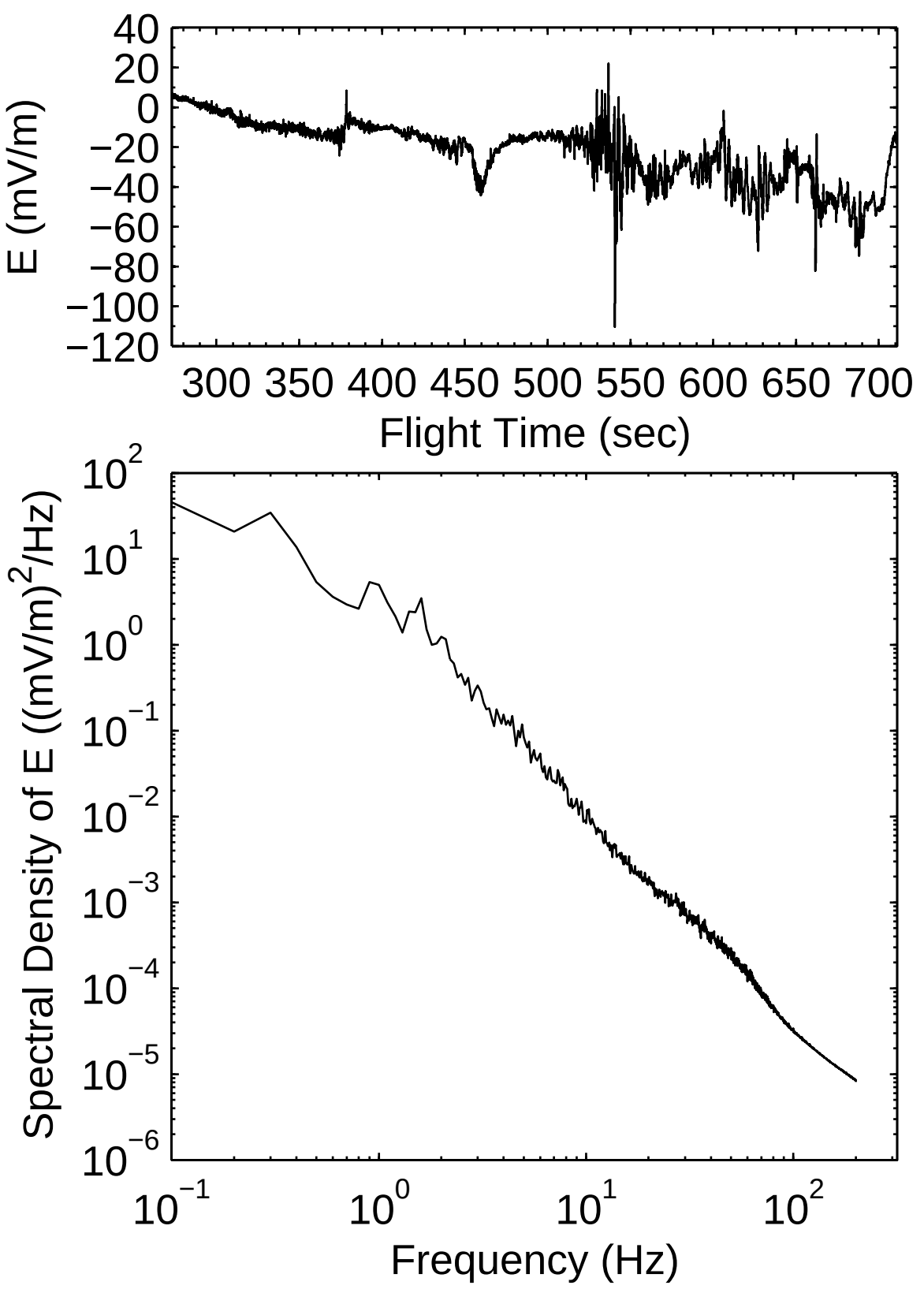

40
20
0
−20
−40
−60
−80
−100
−120
E (mV/m)
300 350 400 450 500 550 600 650 700
Flight Time (sec)
10^2
10^1
10^0
10^−1
10^−2
10^−3
10^−4
10^−5
10^−6
Spectral Density of E ((mV/m)^2/Hz)
10^−1
10^0
10^1
10^2
Frequency (Hz)

Figure 1

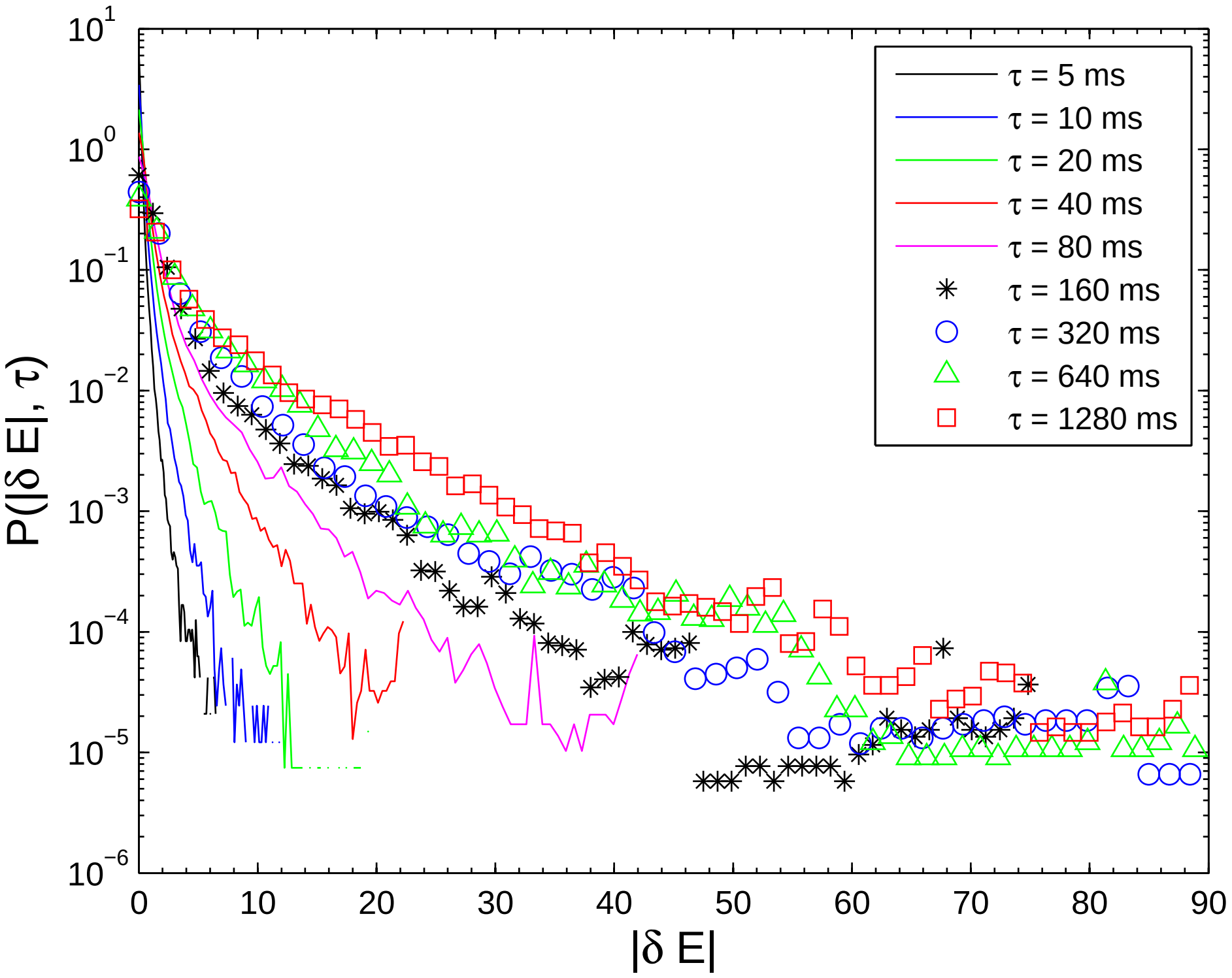

τ = 5 ms
τ = 10 ms
τ = 20 ms
τ = 40 ms
τ = 80 ms
τ = 160 ms
τ = 320 ms
τ = 640 ms
τ = 1280 ms
P(|δ E|, τ)
|δ E|

Figure 2

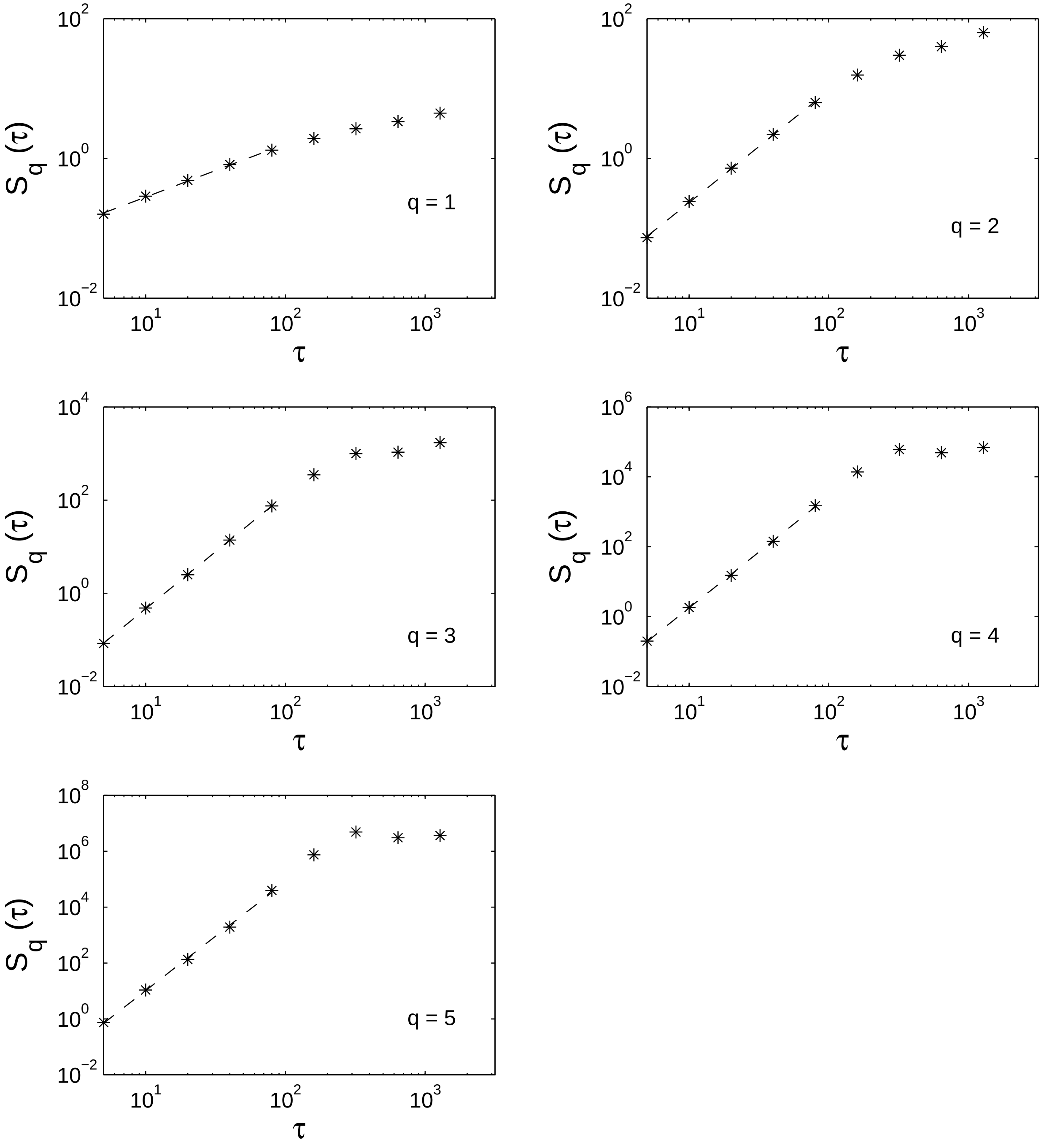

Figure 3

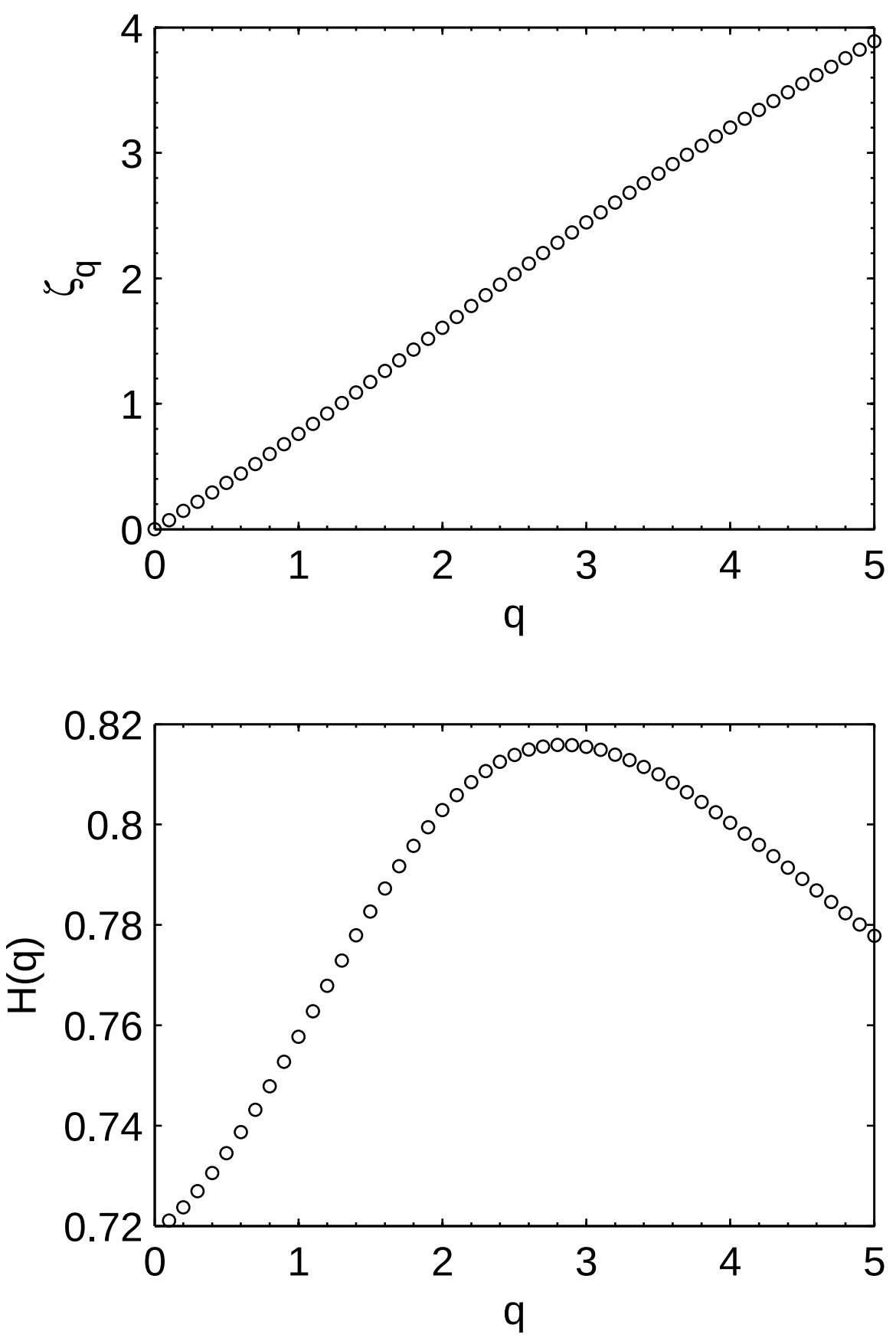

Figure 4

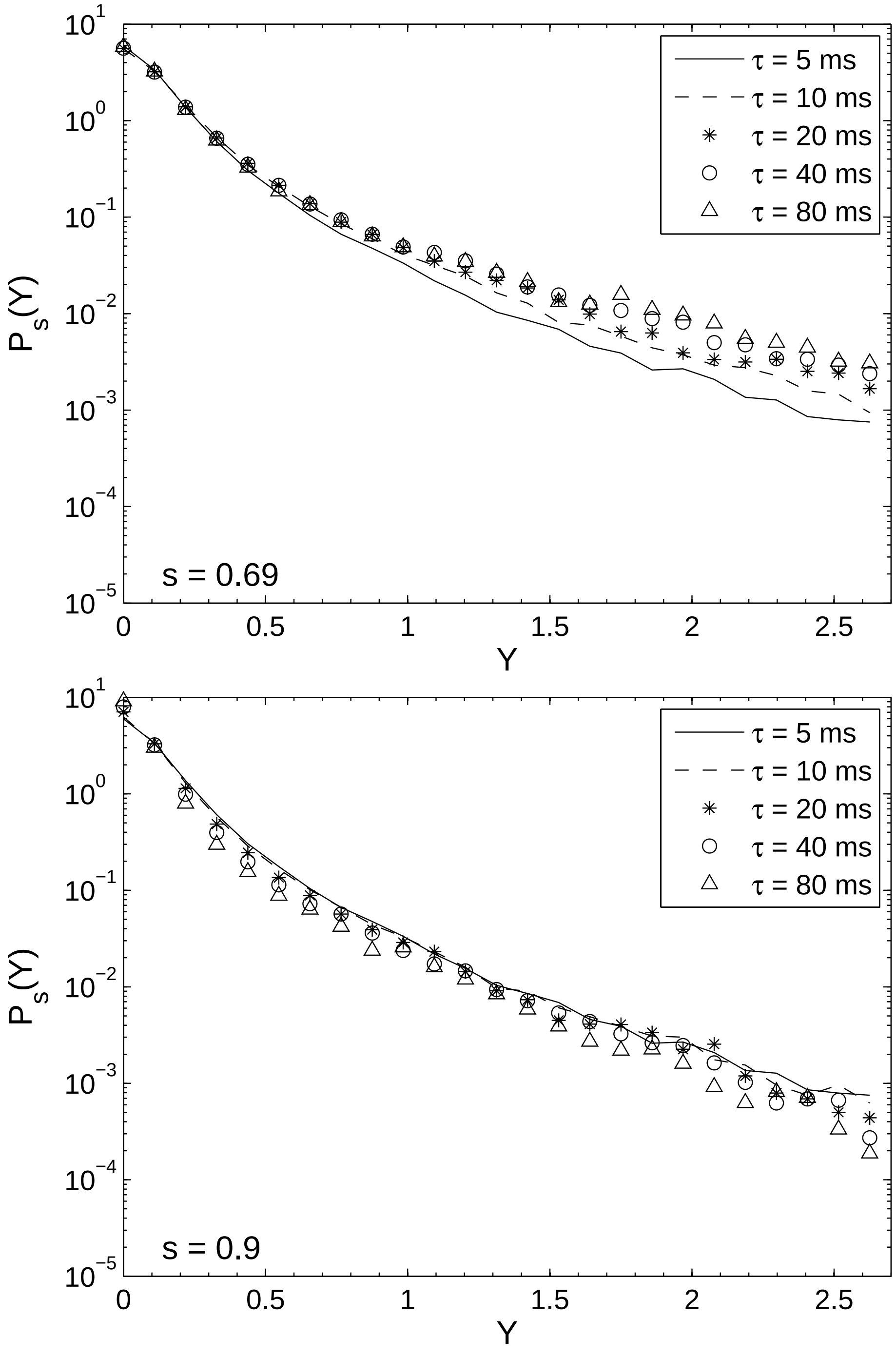

τ = 5 ms
τ = 10 ms
τ = 20 ms
τ = 40 ms
τ = 80 ms
P_s(Y)
Y
s = 0.69
τ = 5 ms
τ = 10 ms
τ = 20 ms
τ = 40 ms
τ = 80 ms
P_s(Y)
Y
s = 0.9

Figure 5

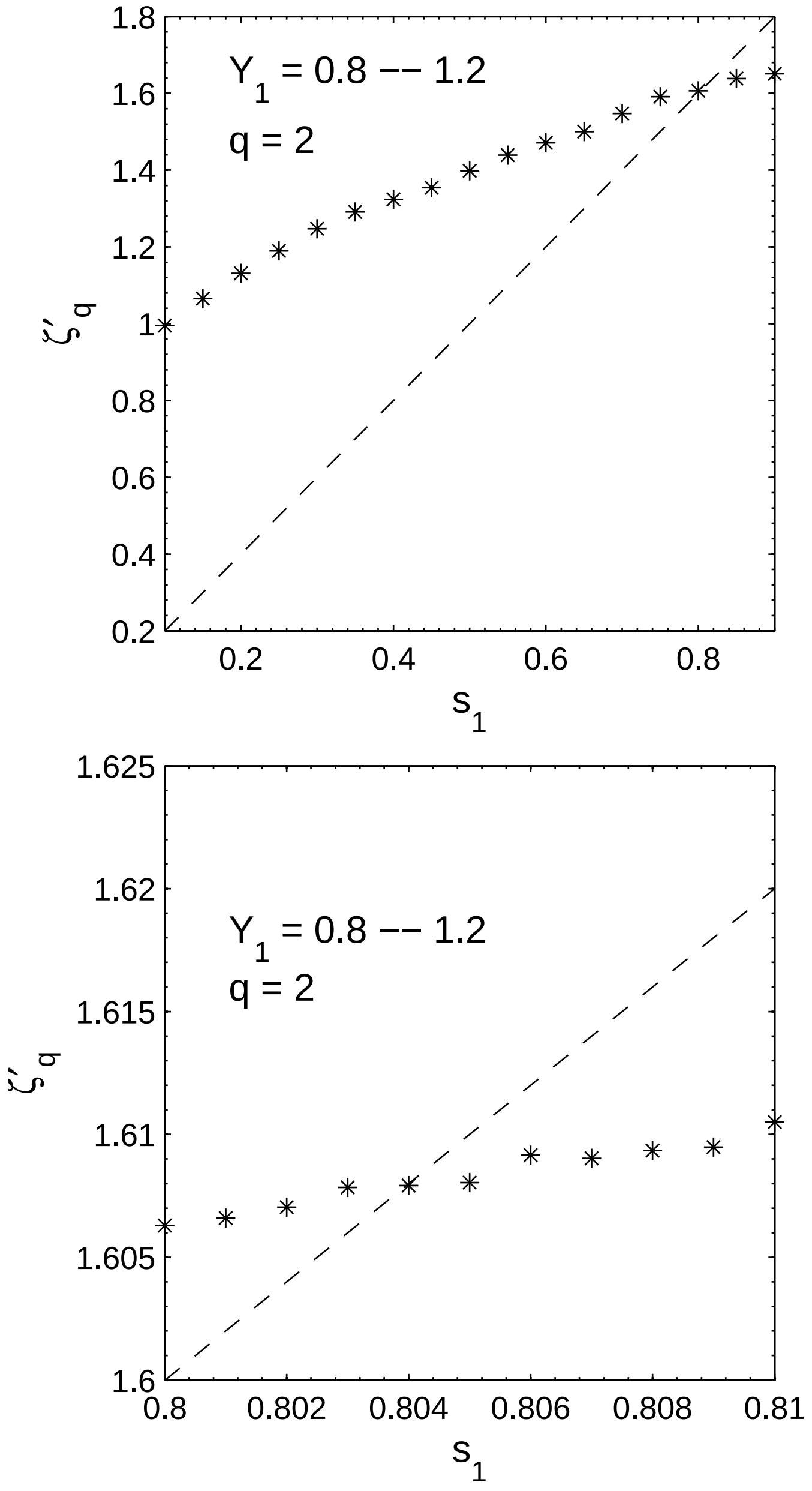

Figure 6

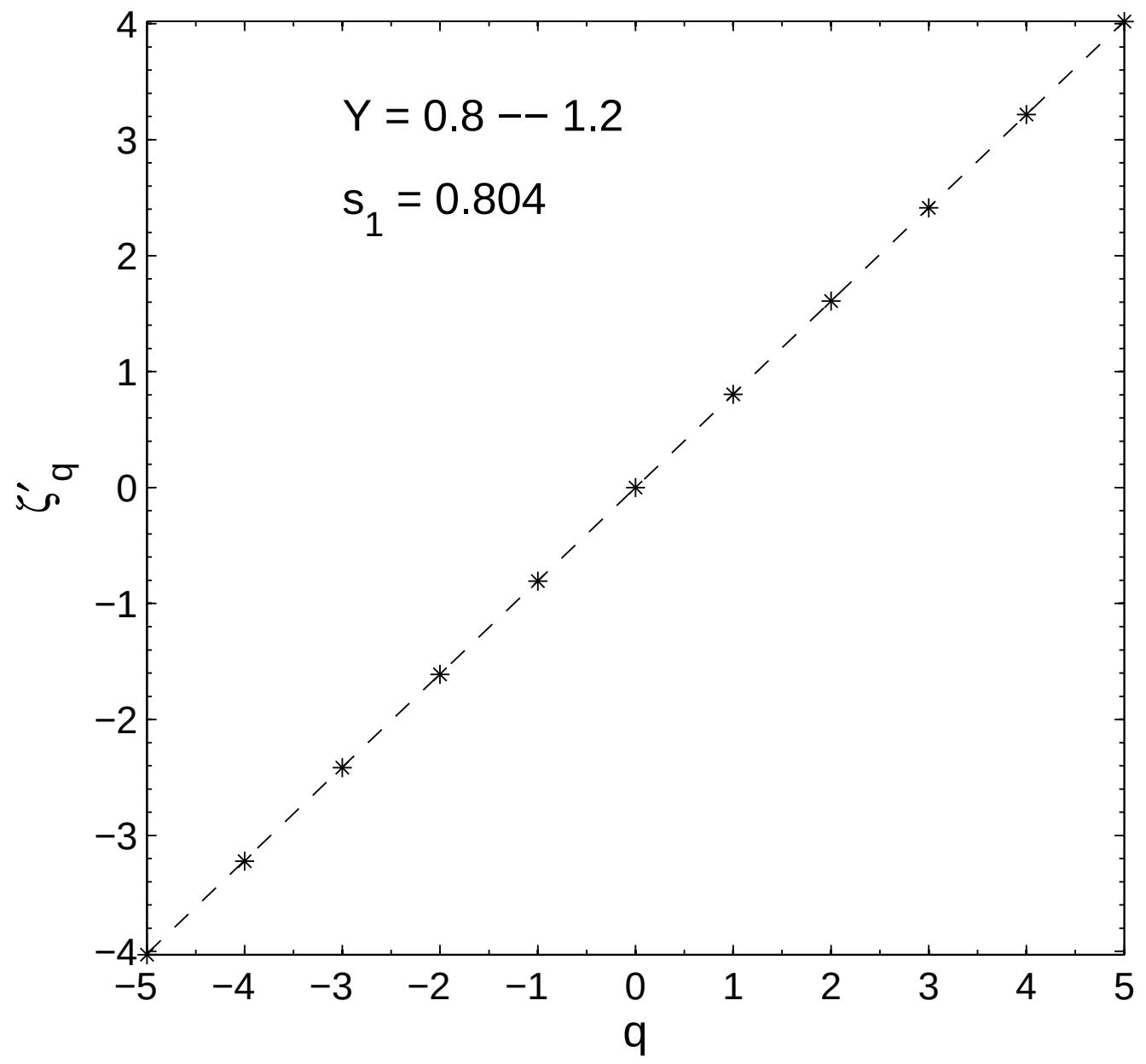

Figure 7

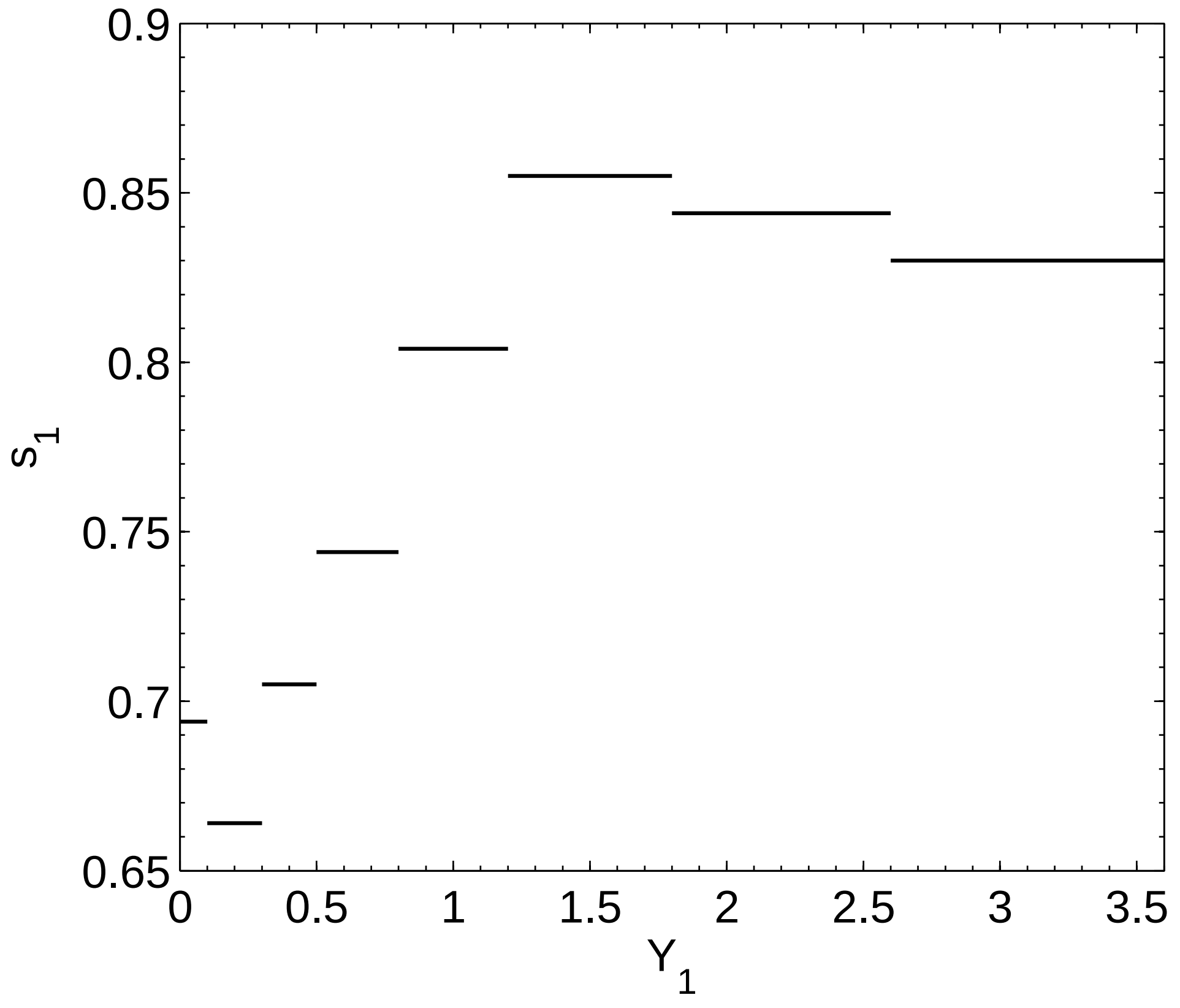

Figure 8

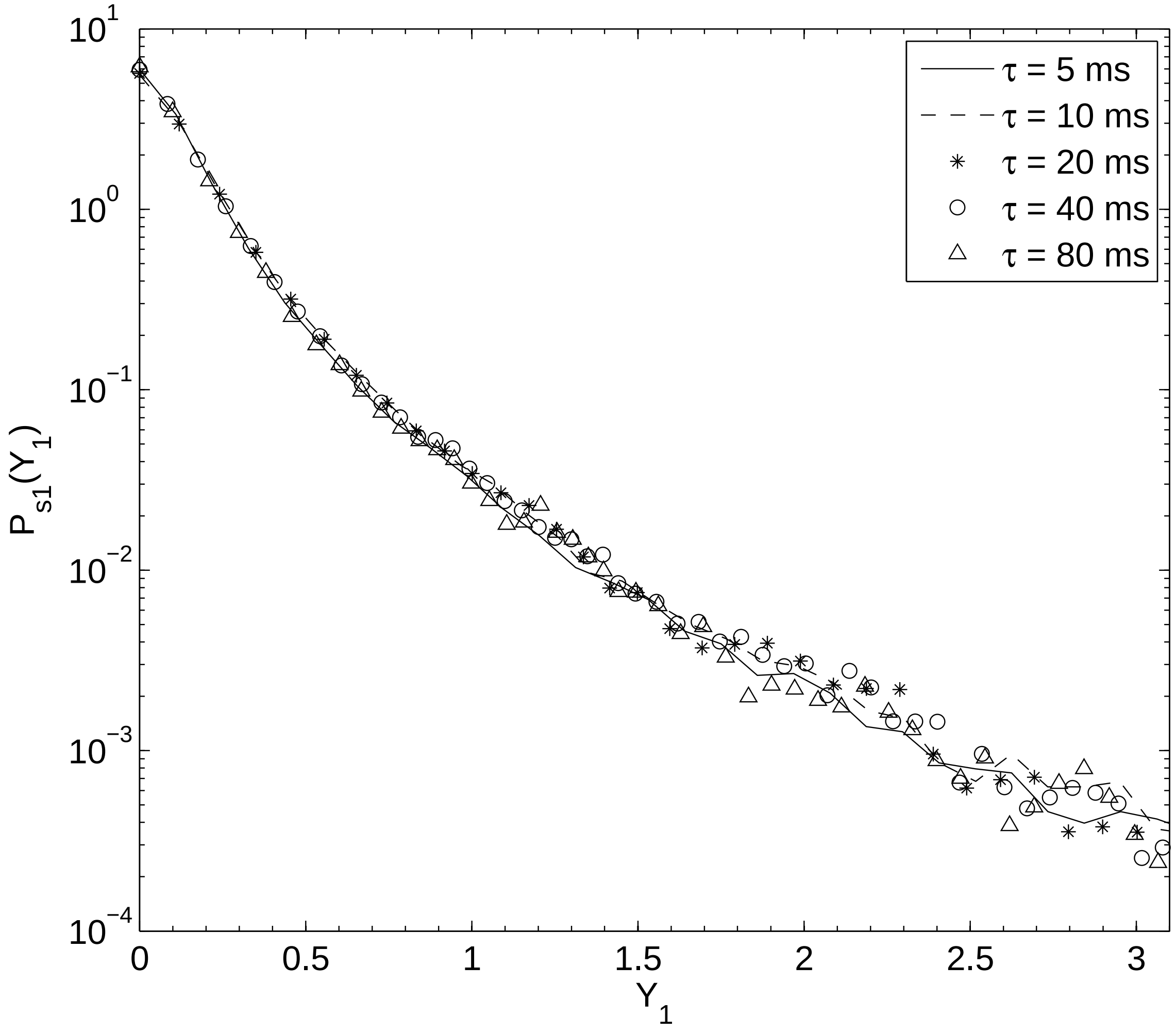
τ = 5 ms
τ = 10 ms
τ = 20 ms
τ = 40 ms
τ = 80 ms
$P_{s1}(Y_1)$
$Y_1$
$10^1$
$10^0$
$10^{-1}$
$10^{-2}$
$10^{-3}$
$10^{-4}$
0
0.5
1
1.5
2
2.5
3

Figure 9

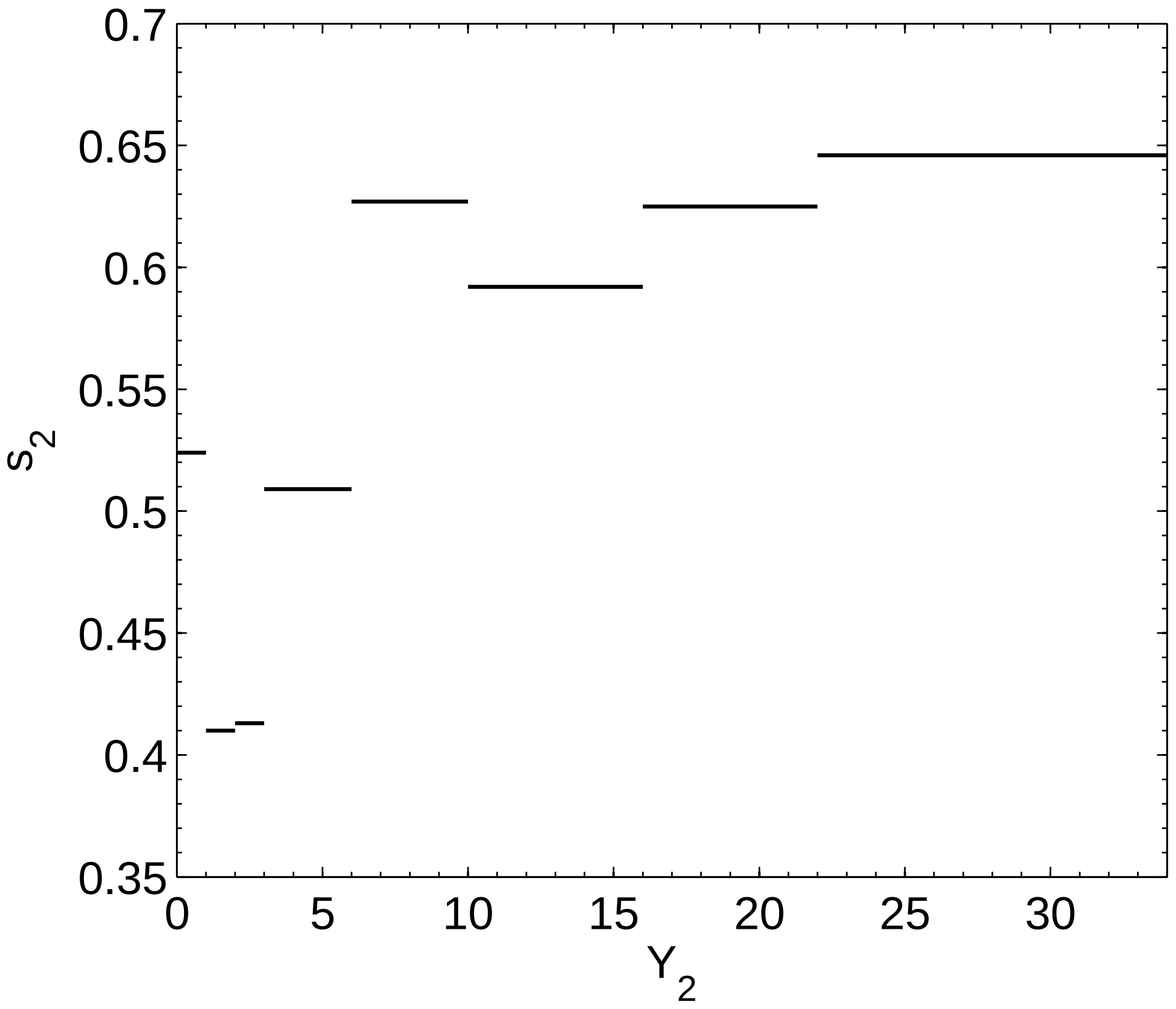

Figure 10

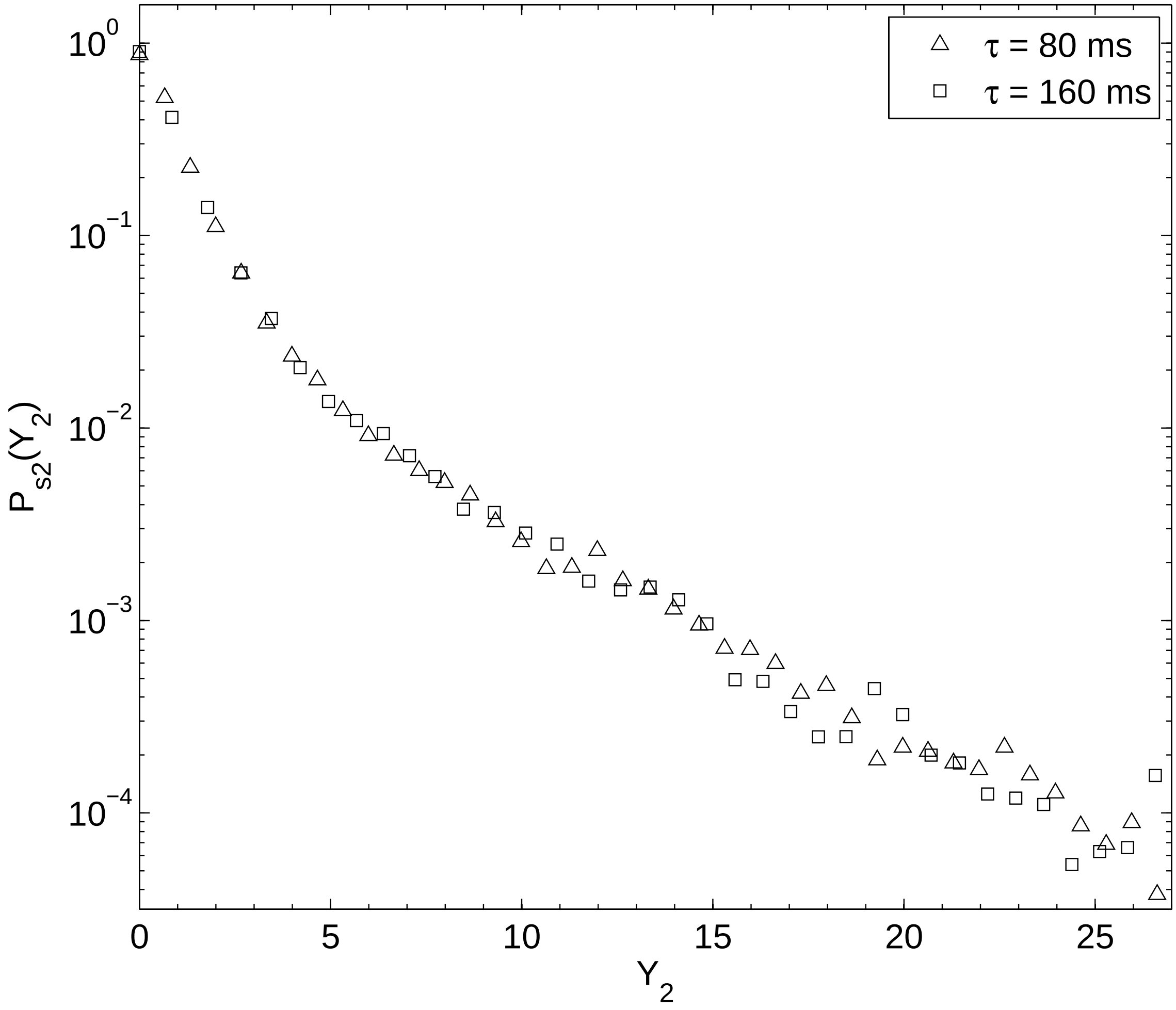

Figure 11

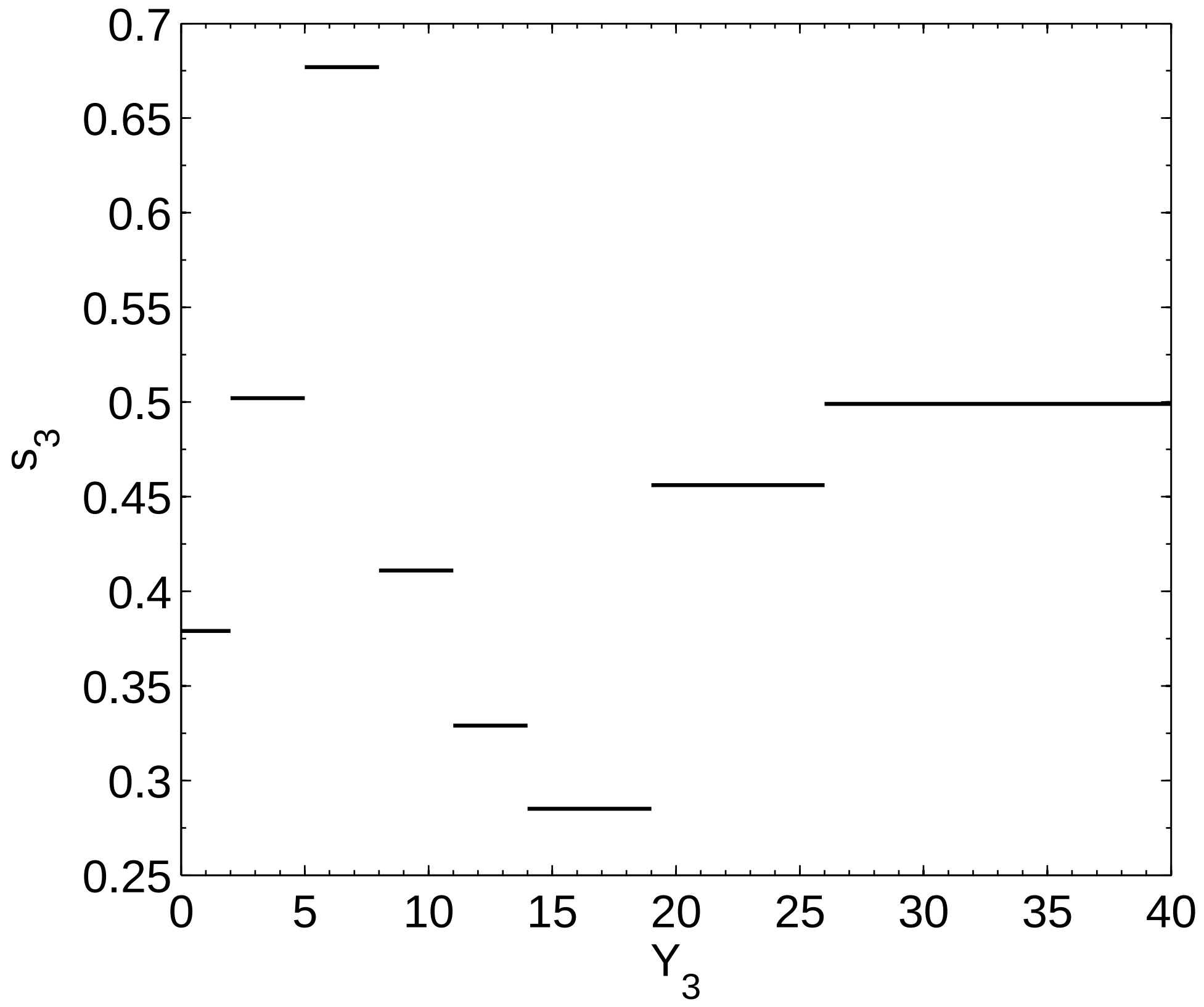

Figure 12

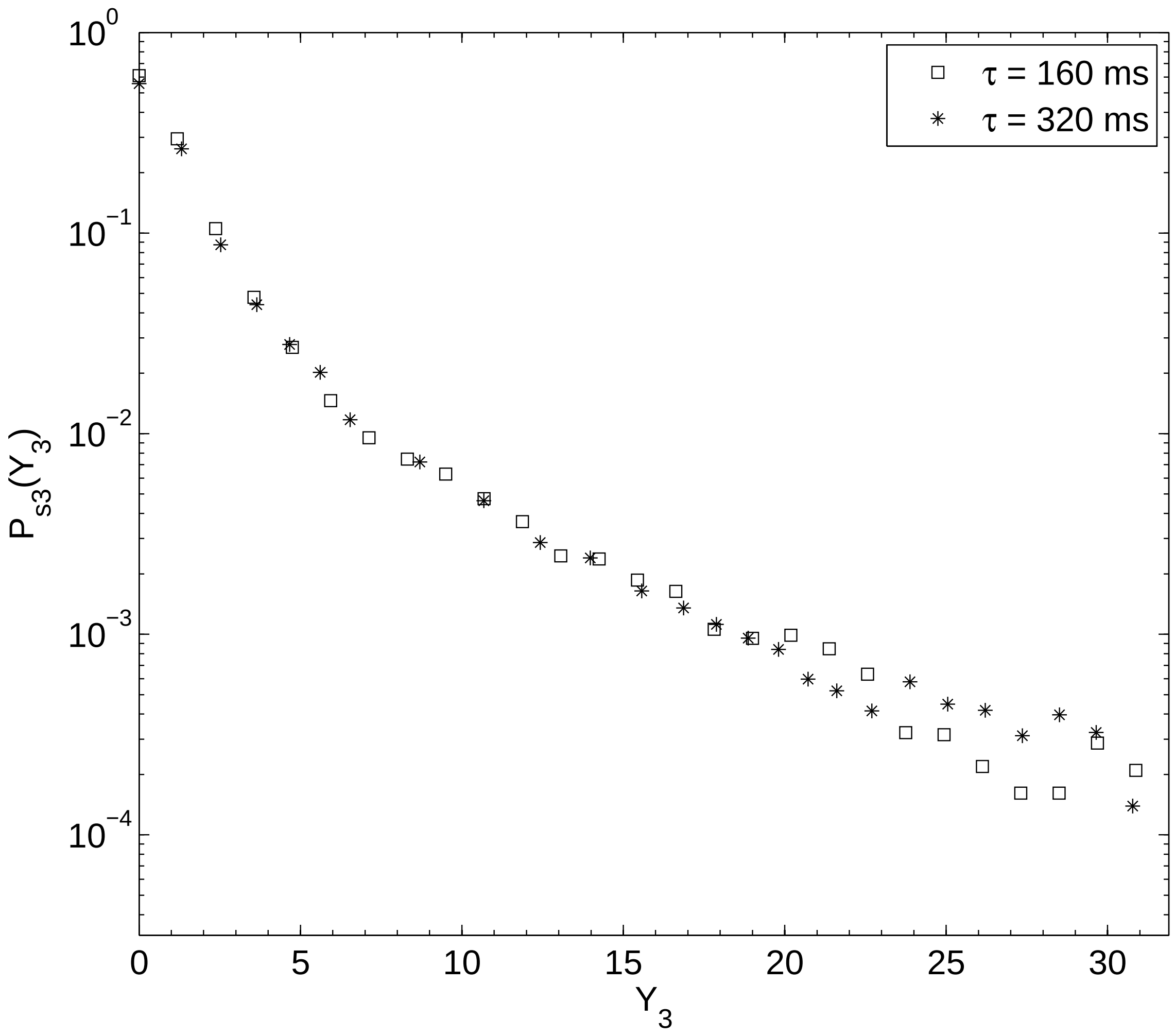

Figure 13

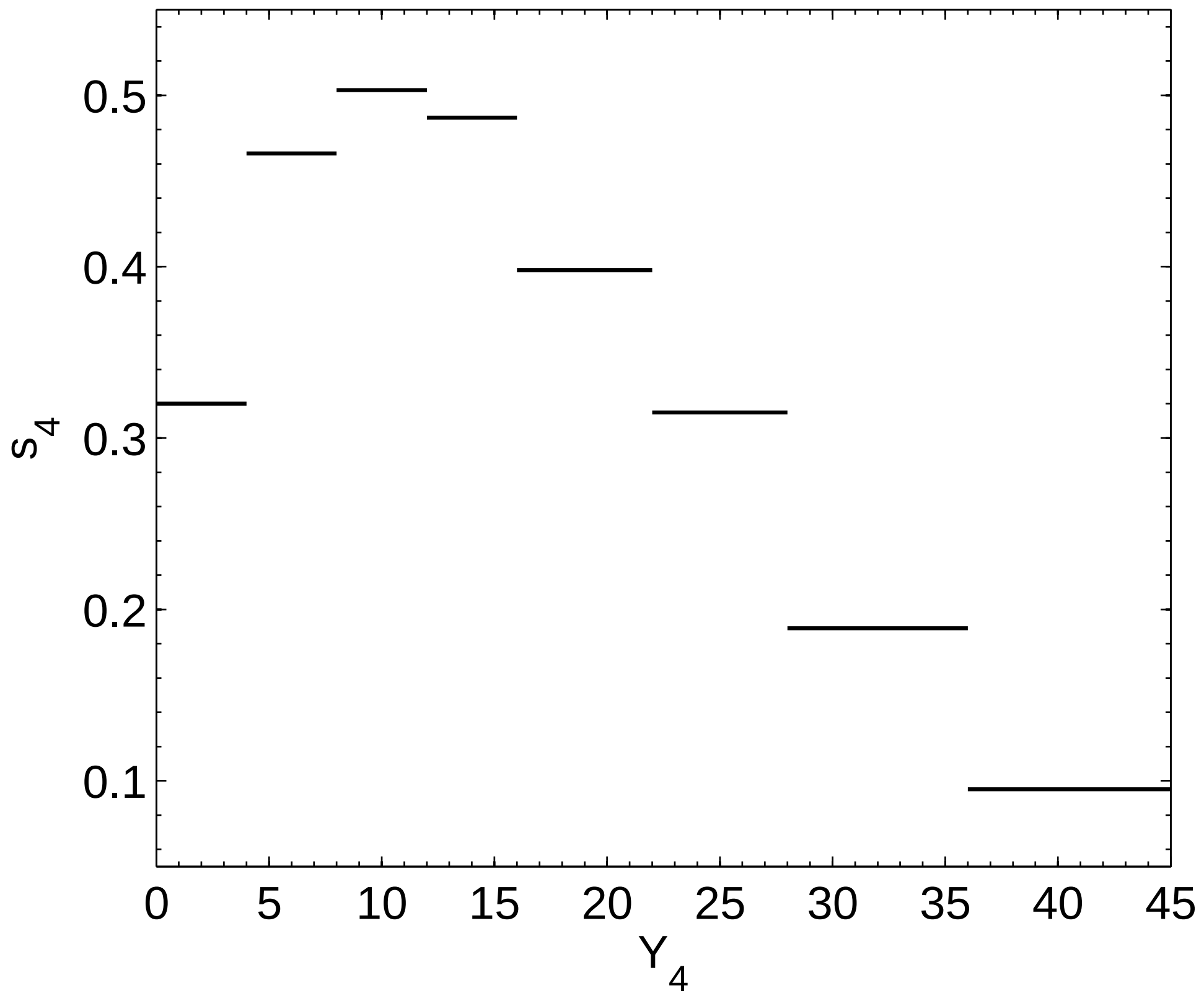

Figure 14

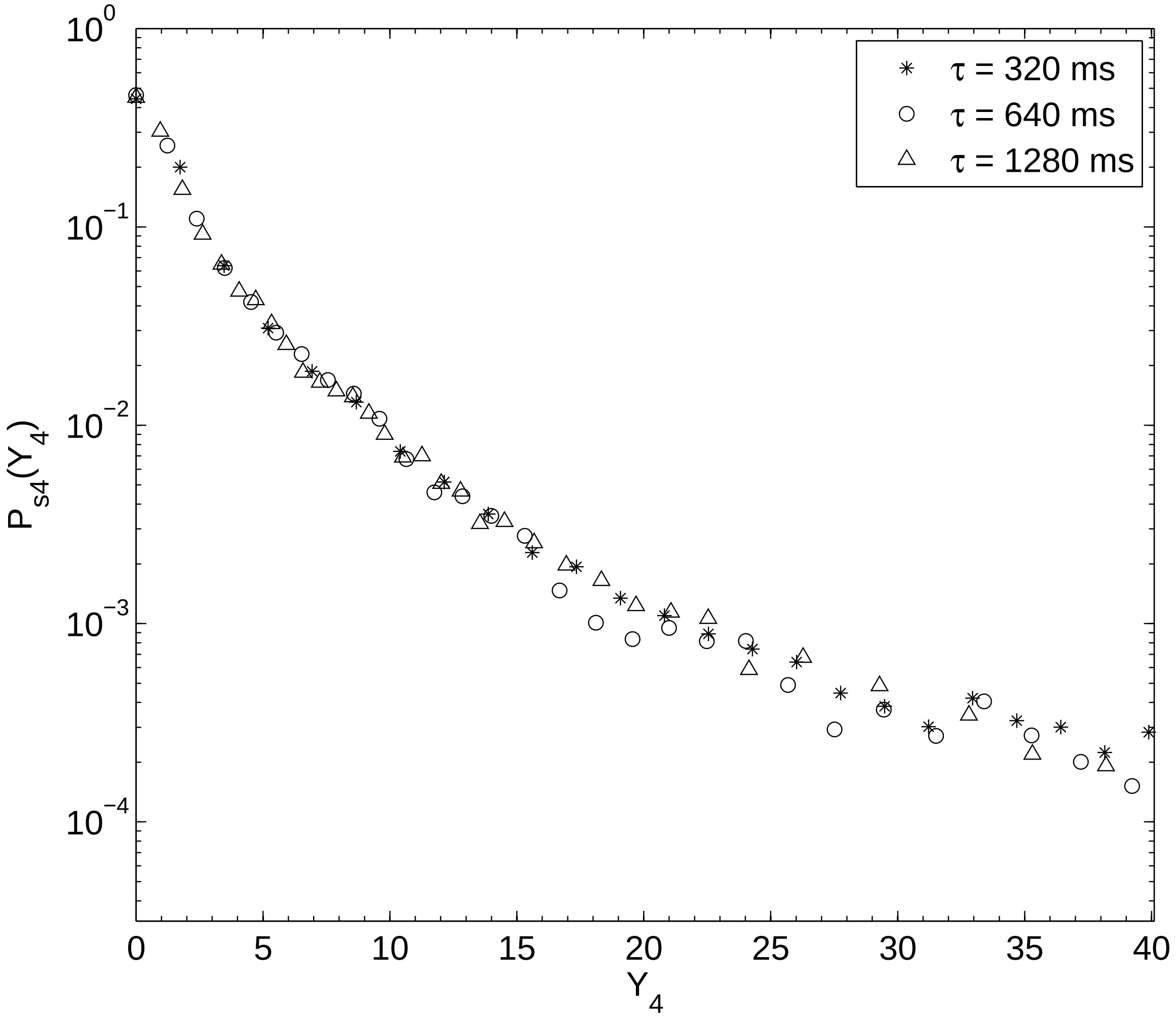

Figure 15

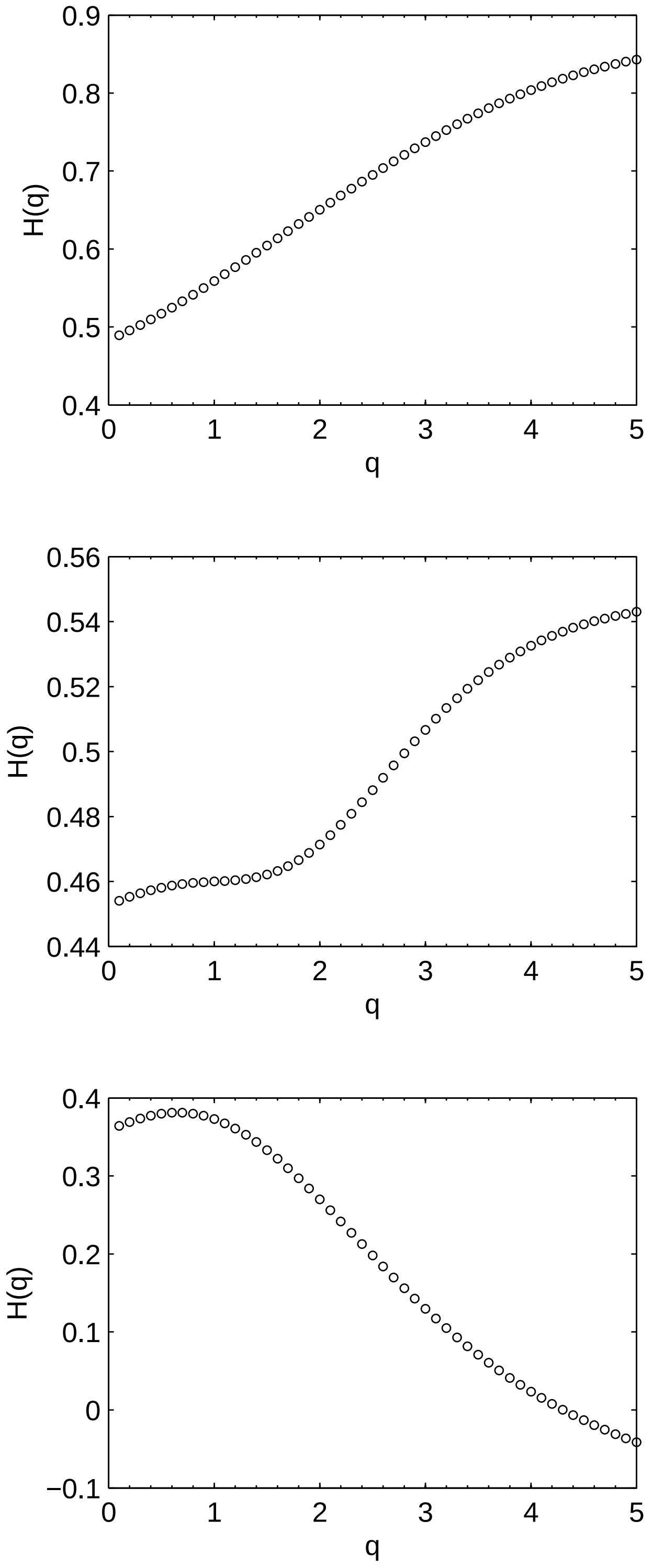

Figure 16

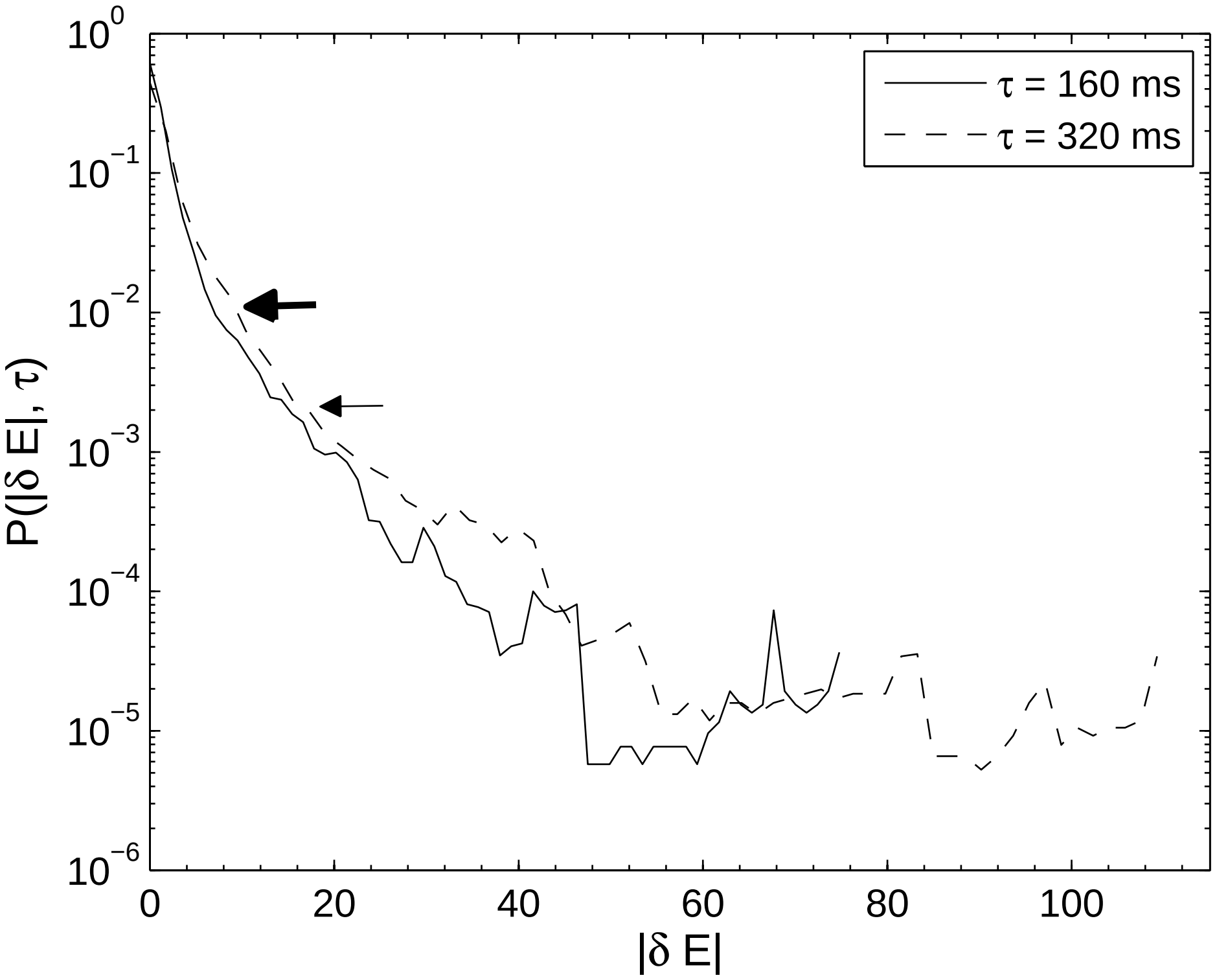

10^0
10^−1
10^−2
10^−3
10^−4
10^−5
10^−6
P(|δ E|, τ)
0
20
40
60
80
100
|δ E|
τ = 160 ms
τ = 320 ms

Figure 17

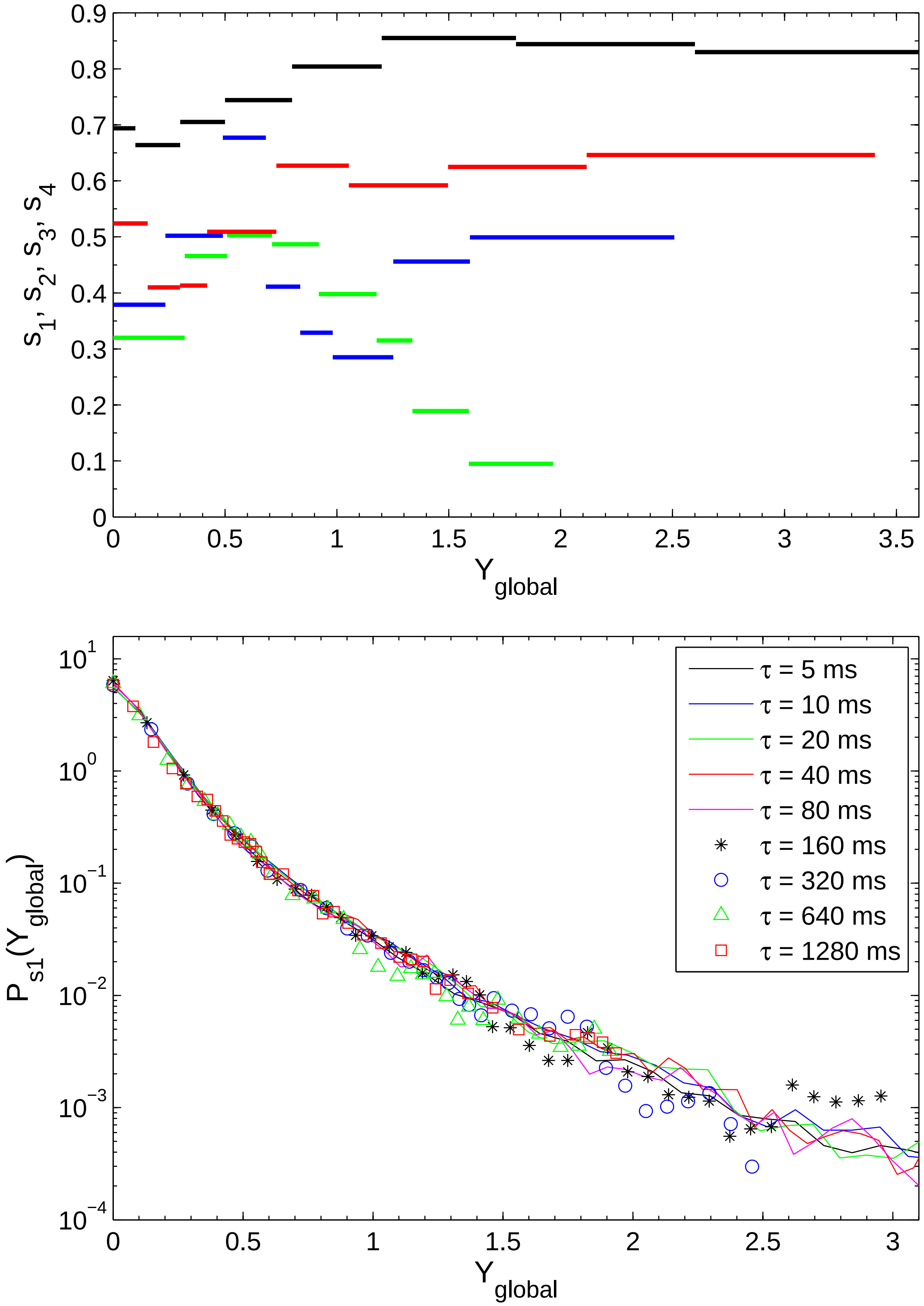

0.9
0.8
0.7
0.6
0.5
0.4
0.3
0.2
0.1
0
s1, s2, s3, s4
0
0.5
1
1.5
2
2.5
3
3.5
Yglobal
10^1
10^0
10^-1
10^-2
10^-3
10^-4
Ps1(Yglobal)
τ = 5 ms
τ = 10 ms
τ = 20 ms
τ = 40 ms
τ = 80 ms
τ = 160 ms
τ = 320 ms
τ = 640 ms
τ = 1280 ms
0
0.5
1
1.5
2
2.5
3
Yglobal

Figure 18